**LED-based multibeam photoacoustics combined with electrical circuit-based modeling for the analysis of multispecies mass transport through thin membranes**

Pawel Rochowski

Institute of Experimental Physics, Faculty of Mathematics, Physics and Informatics, University of Gdańsk, Wita Stwosza 57, 80-308 Gdańsk, Poland

Correspondence: pawel.rochowski@ug.edu.pl

**Abstract**

This work develops photoacoustic-based experimental methods for comprehensive characterization of multispecies mass transport from donor compartments to thin-membrane acceptor systems in perfect contact, supported by dedicated mass transfer modeling framework. Multibeam configurations are implemented in photoacoustic setups operating in front-side detection and diffuse-reflection geometries. The setups are calibrated and adjusted prior to measurements by means of transmission-mode photoacoustic experiments conducted under steady-state conditions. Finally, the methodologies were applied to a model system undergoing photoinduced decay, enabling characterization of bulk transport kinetics as well as interfacial equilibration monitored through time-dependent changes in interfacial reflectivity. For the analysis of bulk transport data, a lumped electrical-circuit (EC) model is introduced. The model is formulated in tableau form, linking species population dynamics to an interaction matrix representing mass-transport couplings consistent with the underlying diffusion–reaction framework. A simplified approximation of the model is further proposed and validated against experimental results. The combined experimental–modeling framework provides an effective approach for quantitative analysis of coupled diffusion, reaction, and interfacial processes in thin-membrane systems.

## 1. Introduction

Studies of transport processes, including mass transport, are central to numerous scientific disciplines, ranging from biology and pharmaceutics to physics and Earth sciences, in both fundamental and applied research [1–4]. A key driver of progress in this area is the common structure of the governing physical laws, which are rooted in the thermodynamic relationship between fluxes and their conjugate driving forces [5]. This general framework underlies the transport of energy, mass, and charge and facilitates a continuous exchange of concepts between disciplines that may otherwise appear only weakly connected at the theoretical and practical levels.

The investigation of mass transport processes inherently relies on appropriate experimental techniques that enable the testing of proposed hypotheses. In applied research, well-established and widely accessible experimental methods, combined with robust data analysis approaches, play a dominant role. A representative example is provided by techniques used in pharmaceutical sciences, such as Franz diffusion cells, which allow the determination of mass transport kinetics across a membrane within the framework of an assumed transport model (i.e. diffusion-controlled transport in the absence of external driving stimuli) [6]. Conceptually similar approaches are employed in column experiments commonly used in Earth sciences [7].

From the perspective of quantitative analysis, it is important to determine not only the amount of substance transported into the medium but also its chemical stability and the impact of possible transformations on the overall transport process. Such capabilities can be provided by spectroscopic detection methods, provided that the investigated species exhibit distinguishable spectral signatures. Another critical aspect is the ability to monitor processes *in situ*, rather than inferring chemical degradation or transformation solely from their indirect effects on release kinetics measured downstream of the membrane. This requirement is addressed by several experimental approaches with varying spatial and temporal resolution, including tape stripping combined with appropriate detection techniques (spectroscopic or radiolabeling), photothermal methods, and, more recently, infrared microspectroscopy combined with Bayesian inference [8–11].

Among the aforementioned approaches, photothermal methods, including photoacoustics, have emerged as powerful tools for non-destructive investigations with high spatial resolution. Their capability arises from the intrinsic mechanism of signal generation, which depends simultaneously on the optical absorption, thermal properties, and structural characteristics of the sample [12,13]. By probing the propagation of thermal waves generated through modulated light absorption, depth-resolved profiling becomes feasible [14,15]. This enables high-precision quantification of mass uptake and, through appropriate modulation schemes, allows monitoring of transport processes as a function of sample depth.

Photoacoustic techniques have long been applied to studies of substance release and mass transport. To date, investigations have been performed in various configurations, employing either modulated continuous-wave excitation or pulsed illumination across spectral ranges extending from the ultraviolet to the infrared. These approaches have consistently provided valuable datasets characterizing transport phenomena. Depending on the assumed transport model, the extracted parameters have included mass diffusion coefficients, thermal relaxation times, and, indirectly, chemical reaction rate constants, binding kinetics, or advective transport contributions [16–19]. The indirect nature of some of these determinations arises from an inherent methodological limitation, namely the monitoring of a single spectral channel during the experiment. This constraint can complicate data interpretation and introduce dependence on the specific transport scenario assumed in the mathematical model.

A possible route to overcome this limitation is provided by multibeam photoacoustics. According to Duhamel’s theorem, the thermal response of a sample to multiple excitations is expected to be linear and additive. Furthermore, the Rosencwaig–Gersho model predicts a linear relationship between the thermally induced piston motion and the acoustic signal detected by the sensor. These premises led to the development of the multibeam photoacoustic (MBPA) modality, in which two or more beams simultaneously probe distinct spectral bands, for example corresponding to different absorbing species within a single sample. The method has been experimentally tested using linear configurations based on low-power, spatially divergent light sources (e.g., LEDs) and a non-resonant photoacoustic cell [20]. The measurements confirmed that the acoustic responses associated with individual excitation beams remain independent, provided that the separation between their modulation frequencies (*Δf*) exceeds both the instrumental frequency instability and the effective detection bandwidth determined by the lock-in time constant and filter characteristics.

These considerations define the objectives of the present work. First, we introduce multibeam photoacoustics for the investigation of complex mass transport in systems containing pigments with distinct spectral properties, enabling direct characterization of multispecies transport without the need for additional ad hoc assumptions within the mathematical framework. In addition, a diffuse-reflectance photoacoustic configuration is implemented for the first time to monitor interfacial properties during the permeation process, providing sensitive indicators of interphase processes occurring at the donor–acceptor and acceptor–air boundaries. Second, the previously developed electrical circuit-based (EC) model of mass transport is extended to incorporate chemical degradation processes, leading to a more general diffusion–reaction EC model.

To validate the proposed methodology and model, an experiment was conducted in which the diffusion of a pigment was monitored between a donor compartment and an acceptor in the form of a thin porous membrane. During the process, the pigment underwent chemical degradation, forming a secondary species that was detected simultaneously using an additional probe beam. As a result, two independent release profiles were obtained, enabling the determination of diffusion coefficients and decay rates separately in the donor and acceptor domains. Simplified versions of the model were also introduced to allow straightforward estimation of the key transport parameters. Finally, characteristic times associated with interfacial equilibration were determined based on photoacoustically detected changes in the reflectivity of the donor–acceptor and acceptor–air interfaces.

## 2. Theoretical background

### 2.1 The passive mass transport

For the purpose of the work, we consider a simple 1D transient system consisting of two compartments in contact, namely a mass donor (*A*) and a mass acceptor (*B*), and assume $A \rightarrow B$ mass transport direction. It is assumed that the mass transport is passive (i.e. in the absence of any external stimuli) and obeys the Fick's laws of diffusion:

$$j = -D\frac{\partial}{\partial x}c \tag{1a}$$

$$\frac{\partial c}{dt} = D\frac{\partial^2}{\partial x^2}c \tag{1b}$$

Eq. 1a gives the mass flux $j$ as proportional to the concentration gradient $\frac{\partial}{\partial x}c$, with the proportionality factor – the diffusion coefficient $D$. We consider the concentration as $c(x,t) \equiv c = n/l$, where $n$ – number of trace particles, $l$ – the reduced (1D) volume or, equivalently characteristic length. Eq.1b is valid for homogeneous systems ($D$ is constant) and describes the mass accumulation/depletion rate as proportional to the local curvature of the concentration gradient.

The analysis of mass transport behaviour primarily depends on the type of experimental data considered. In general, experimental modalities provide mass permeation data in two forms. The first involves *c(x,t)*-type data, i.e., the temporal evolution of concentration distributions. In this case, the analysis relies on a specific mathematical expression for *c(x,t)* (the solution to Eq. 1b), which depends on system properties such as geometry or symmetry, boundary and initial conditions defined by the experimental setup, as well as additional physicochemical parameters of the interfaces affecting intercompartmental mass transfer kinetics.

The second form consists of *n(t)*-type data, i.e., cumulative sorption profiles, analysed using the relation: $n(t) = \int c(x,t)dx$ where the integration limits span the domain of interest. A detailed discussion of the applicability and reliability of *c(x,t)*- and *n(t)*-type data for mass transfer quantification is provided in [11]. In the present work, we aim for the lumped *n(t)* approach or its equivalent spatially averaged *c(t)* formulation.

Recently, a novel approach for quantifying *n(t)*-type data and mass transfer in composite systems has been proposed. The model exploits the formal analogy between mass and charge transport laws (Fick's and Ohm's laws), establishing pairs of equivalences such as mass–charge, volume–capacitance, and concentration difference–voltage [21]. By applying Kirchhoff's rules, it enables prediction of mass transfer rates between subsystems in perfect contact. For a simple $A \rightarrow B$ case, the mass evolutions can be derived from the evolution equation for $n_A(t)$ ($\equiv n_A$) and $n_B(t)$ ($\equiv n_B$):

$$\frac{d}{dt}\begin{pmatrix} n_A \\ n_B \end{pmatrix} = \begin{pmatrix} -\frac{1}{RV_A} & \frac{1}{RV_B} \\ \frac{1}{RV_A} & -\frac{1}{RV_B} \end{pmatrix}\begin{pmatrix} n_A \\ n_B \end{pmatrix} = \begin{pmatrix} -\Sigma_A & \Sigma_B \\ \Sigma_A & -\Sigma_B \end{pmatrix}\begin{pmatrix} n_A \\ n_B \end{pmatrix} \tag{2}$$

where $R = \frac{l_B}{D^{EC}}$ is the $B$ material (the acceptor) resistivity, $V$ stands for the compartmental ($A$ or $B$) volume and $\Sigma = \frac{1}{RV}$. Solution of the Eq. 2 requires specifying the initial conditions. As an example, by setting the $n_A(0) = 1$, $n_B(0) = 0$ (all the mass initially in the acceptor compartment) and considering the reduced volumes of *A* and *B* compartments as $l_A$ and $l_B$, the solution for $c_B = n_B/l_B$ takes the exponential form:

$$c_B = \bar{c}(1 - \exp[-\Sigma t]) = \bar{c}(1 - \exp[-t/\tau]), \tag{3}$$

where $\bar{c}$ is the average mass concentration in the system (in general, $\sum_i n_i \,/\, \sum_i l_i$), $\tau = RV = \frac{l_B}{D^{EC}} l_{eq}$ stands for the characteristic time of the process and $l_{eq}$ is the equivalent volume ($l_{eq}^{-1} = l_A^{-1} + l_B^{-1}$). The solution is reminiscent of the single RC circuit voltage charging curve. It is understood that the $c_A$ evolution follows the same kinetic pattern as $c_B$ (due to the mass conservation principle, the mass loss in *A* is compensated by the mass rise in *B*).

The EC model as stated does not refer to any specific geometry. As such, it is crucial to add correction factor to the estimated diffusion coefficient to meet desired convergence to the Fick's model predictions. In the case of the plane sheet geometries, the relationship between the EC-diffusion coefficient $D^{EC}$ and the actual diffusion coefficient $D$ is: $D^{EC} = 2.62D$.

Due to its simplicity, flexibility and reasonable level of precision, the EC approach is considered as a promising tool for analysing the mass transport in composite and multispecies systems. Section 4 outlines an extension of the approach to the A → B system incorporating decay.

### 2.2 Photoacoustics

Photoacoustic methods are based upon the detection and analysis of *the photoacoustic effect* (PA), that is generation of pressure waves followed by the absorption of modulated energy flux by a sample of thickness $l_s$. In a typical experimental configuration as explored here, i.e. a front-side microphone detection (with air as an intermediary medium) of the PA effect, the 1D signal generation can be described by the theory developed by Rosencwaig and Gersho (the RG model). In general, the authors assumed that the PA signal generation is a three-step process, consisting of:

- perturbation of the temperature field in a sample, followed by the Lambert-Beer type of absorption of modulated light. The heat source distribution is given by:

$$h(x,t) \propto A\,\eta\,I\,exp[-\alpha x] \propto \frac{1}{2}\,A\,\eta\,I_0(1+\cos[\omega t])\exp[-\alpha x], \qquad (4)$$

where: $A$ is the apparatus constant, $\eta$ stands for the light-to-heat (thermal, non-radiative deexcitation) energy conversion coefficient, $\omega$ is the modulation frequency in angular units ($\omega = 2\,\pi f$) and $\alpha$ is for the light absorption coefficient;

- evolution of the temperature field obeying set of Fourier-Kirchhoff equations for the backing (subscript *b*), sample (s) and gas (g), with a proper continuity conditions between the domains:

$$\frac{1}{\beta_{b,g}}\frac{\partial T}{\partial t} = \frac{\partial^2 T}{\partial x^2} \qquad \text{for the } b \text{ and for the } g \text{ domains,} \qquad (5)$$

$$\frac{1}{\beta_s}\frac{\partial T}{\partial t} = \frac{\partial^2 T}{\partial x^2} + \frac{h(x,t)}{k_s} \qquad \text{for the sample domain,} \qquad (6)$$

where β stands for the thermal diffusivity, while κ is for the thermal conductivity;

- adiabatic energy transfer from the excited sample to the gas phase, and the following periodic gas pressure variations $\delta p$ (due to the thermal piston action of a gas layer adjanced to the sample) recorded by a detector. The PA signal *S*:

$$S \propto \delta p \equiv Re\left[Qexp\left[i\left(\omega t - \frac{\pi}{4}\right)\right]\right] = q\cos[\omega t - \psi - \frac{\pi}{4}] \qquad (7)$$

where $Q$ stands for complex envelope of the gas pressure variations, while $q$ and $\psi$ specify the magnitude and phase of $Q$. Eventually, the PA signal $S$ depends on all of the system thermal and optical characteristics encoded in $Q$: $S \equiv S(b,s,g,f^{-n})$, where $n \in (1;1.5)$ and depends on an interplay between the geometrical, optical and thermal properties of the sample.

Photoacoustics method can be employed for the material characterization in two distinct ways. When measured against varying wavelength (Λ) and constant modulation frequency (PA spectroscopy mode), in the absence of saturation effects, the PA response is proportional to $S(\Lambda) \propto AI(\Lambda)\eta(\Lambda)\alpha(\Lambda)$, and exhibit similarities to the standard absorption spectroscopy data (the deviations originate from the presence of $\eta(\Lambda)$ term). When measured against varying modulation frequency and constant wavelength (PA depth-profiling mode), the signal is proportional to the $S(\omega) \propto A\eta f^{-n}$. An advantageous character of the method appears for the measurements in the thermally-thick region, that is when the thermal diffusion depth, controlled by the modulation frequency, drops below the sample thickness. Then, the PA sampling depth (i.e. a layer contributing to the PA signal formation) can be approximated by $L(f) \equiv L \approx \sqrt{\frac{\beta_s}{\pi f}}$. By considering the PA magnitude proportional to the pigment concentration, it is possible to measure total amount of pigment within the sample layer of interest ($m(t) = \int_0^L c(x,t)dx$) as long as the $1/\alpha > L$ relation holds (i.e. optically quasi-transparent regime of the RG approach).

### 2.3 Multibeam (MBPA) and diffuse reflection (DRPA) photoacoustics

The photoacoustic multi-beam modality involves multiple scanning beams applied simultaneously for the sample monitoring. In the case of divergent and relatively weak light sources (like LEDs for instance), i.e. where all the nonlinear effects are negligible, the cumulative sample PA response is given simply by a sum of individual contributions to each excitation. If the excitations are characterized by various modulation frequencies, then the lock-in process allows for an unambiguous differentiation of the contributions as long as the frequency shift between the modulations exceeds the filter bandwidth.

A basic two-beam setup for the *f*-domain MBPA in the front-side geometry (FSD) detection mode, as extensively reported in [20], utilizes two separate (by means of wavelength and modulation frequency) scanning beams routed by a dichroic mirror to a sample sealed in a photoacoustic chamber (Fig. 1a). A cumulative PA signal recorded by a microphone is passed to two lock-in units and analysed with respect to the reference modulation frequency provided by a two-channel function generator driving the LEDs. It is understood that the magnitudes of each contribution depend on the particular excitation wavelength-dependent $\alpha(\Lambda)$ and $\eta(\Lambda)$ (i.e. each spectral site can be characterized by distinct light-to-heat conversion factors), but share the same $\beta_{b,s,g}$.

The MBPA method for the mass transport parameters quantification can be employed in two distinct ways. The first one is based on the PA FSD depth-profiling modality, which allows to probe overall amount of the light-absorbing pigment molecules within a material layer of interest. By utilizing two incident wavelengths it is possible to track the mass uptake of two molecular species of distinct spectral characteristics simultaneously. It is understood that application of two probing beams of the modulation frequencies such that $f_2 = f_1 + \Delta f$ with $\Delta f$ exceeding the filter bandwidth (condition necessary to preserve independence of the responses to the two probing beams in the MBPA configuration) leads to a certain alteration of the sampling depth of the second probe beam with respect to the first one.

By taking the sampling depth of $L \equiv L_1 = \sqrt{\frac{\beta_S}{\pi f_1}}$ and the altered length of $L_A \equiv L_2 = \sqrt{\frac{\beta_S}{\pi(f_1+\Delta f)}}$ one obtains a relation between the separation frequency and the relative sampling depth alteration magnitude:

$$f_1 = \Delta f\left(\left(\frac{L}{L_A}\right)^2 - 1\right)^{-1} \quad (8)$$

which for a given $\Delta f$ provides an estimate of a cut-off frequency, allowing to attain the desired sampling depth accuracy. For example, with 98% accuracy (i.e. $L_A/L \approx 0.98$) and a separation frequency of 1 Hz, the cut-off frequency is ~24 Hz. The condition allows to minimize a mismatch between the sampling depths and the sample thickness.

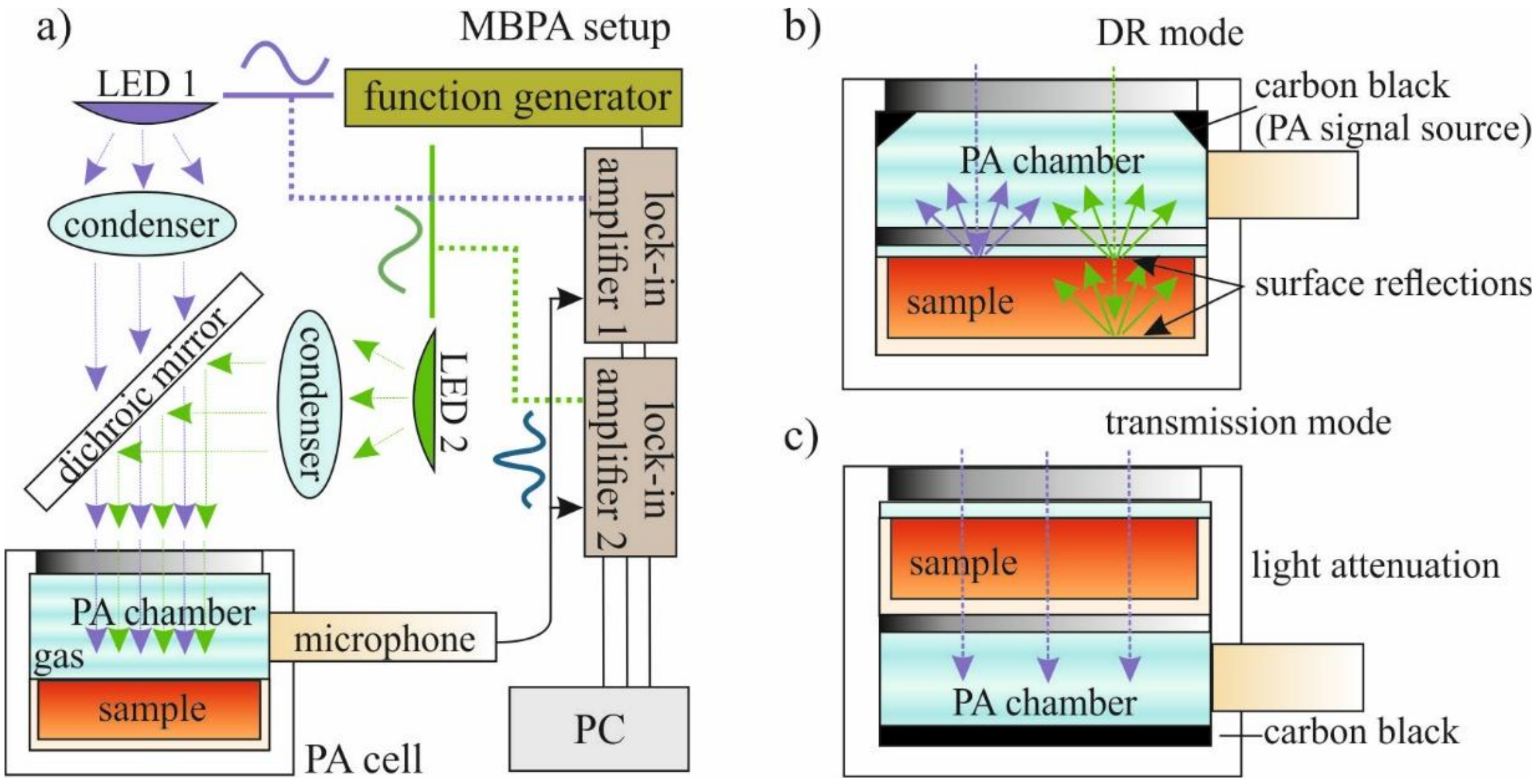

Fig. 1: Block diagrams for: a) the MBPA setup, including the front-side detection scheme for determining the bulk mass transport parameters; b) the PA chamber overview for diffusive reflection studies, examining interfacial processes; c) the PA chamber in transmission mode, estimating sample optical parameters and conducting stationary spectral band analysis for transport experiments.

The second of the considered PA-based modality for the mass transport quantification is based upon the diffusive-reflectance (DRPA) method (Fig. 1b), where one probes the sample surface back-scattered light [22]. The backscattered light can be of two origins: direct, where the incident light is being reflected by a surface or indirect, where the radiative deactivation of the excited pigments takes place. It is understood that the impact of the indirect contribution can be varied and minimized by a proper materials selection or chemical modification of the pigments analysed. Then, the magnitude of the direct contribution can be quantified upon Fresnel model for the reflection and transmission of light. In brief, for a non-polarized light (as in case of LEDs) the effective reflectivity of a material is given by an average of the s- and p-type reflectivities, and depends on the refractive indices ($N$) of the optical media and the angles of incidence and refraction. For the special case of normal incidence, the effective reflectivity can be simplified to $R_{eff}(0^o) = \left(\frac{N_2-N_1}{N_1+N_2}\right)^2$. For an exemplary case of a polymer film (membrane) surrounded by air, $N_1 = 1$ and $N_2 = 1.5$ (typical value for polymers), the power reflectance at normal incidence is only 4%. Finally, the detection limits for DRPA modality are determined by the small amount of the probing light backscattered to the detector chamber. A somewhat more physical approach for the direct DRPA origin involves introduction of the complex magnitude of the refraction index: $\widetilde{N} = N + ik$, where $k\,(= \alpha\Lambda/4\pi)$ quantifies the light absorption in the medium. As the light absorption is generally wavelength-dependent, the recorded backscattered light provides surface absorption patterns, affected by the presence of certain molecular groups.

The third method is considered a supplementary PA modality and involves measurements in transmission mode (Fig. 1c). In this method, a carbon black pastille sealed in a chamber beneath an absorbing sample serves as a PA signal source. This method measures the amount of light passing through the sample, enabling the $\alpha(\Lambda)$ characteristics to be tracked while maintaining the experimental conditions of MBPA and DRPA measurements.

### 2.4 The observables

To relate the experimentally observed photoacoustic signals to the evolution of pigment concentration in the membrane, we refer to the model presented in the follow-up studies presented in [23].

The first fundamental assumption of the model is a clear separation of timescales between thermal processes and mass transport. In particular, assuming that mass transport processes are slow compared to the thermal processes responsible for PA signal generation, the heat source distribution can be considered effectively frozen on the PA timescale. This allows the use of standard formalisms for PA signal generation developed for stationary systems.

The second key assumption follows directly from the nature of mass transport. In the case of diffusive transport considered here, the driving force is the gradient of pigment concentration. Consequently, the observed PA signal evolves from the initial state, corresponding to the pigment concentration at $t = 0$ within the layer $l$, toward a final steady state in which the pigment concentration becomes uniform throughout the system.

#### 2.4.1 FSD technique

In the case of front side detection (FSD), the photoacoustic signal is directly proportional to the AC component of the sample surface temperature, $S_{FSD} \propto |\,\Theta(0)\,|$. Within the Green's function formalism, the surface temperature is given by $\Theta(0) \propto \int_0^l G(0,x')\,h(x')\,dx'$, where $G$ depends on the thermal properties of the system, and $h$ represents the heat source distribution [24]. Thus, temperature variations are determined by the forms of both $G$ and $h$.

For the case considered in this work, as a first approximation, the transport process may be assumed to occur in an optically transparent regime. It can then be described by the model analysed by Crank, i.e., diffusion from a semi-infinite reservoir ($c_A =$ const throughout the process) into a thin finite membrane ($c_B(x, 0) = 0$). In this case, the solution takes the form:

$c_B(x, t_0) = c_A \left(1 - \frac{4}{\pi} \sum_{j=0}^{\infty} \frac{(-1)^j}{2j+1} \exp\left[-\frac{t_0}{\tau_j}\right] \cos\left[k_j x\right]\right)$, where $\tau_j = \frac{4l^2}{D(2j+1)^2\pi^2}$, $k_j = \frac{(2j+1)\pi}{2l}$, and $t_0$ denotes the time parameter defining the (frozen) pigment distribution (Eq. 2.67 in [25]).

In the optically transparent regime, the effectively frozen heat source distribution can be written as $h(x, t_0) = \eta I \varepsilon c_B(x, t_0) \exp\left(-\varepsilon \int_0^l c_B\,(x', t_0)\, dx'\right) \approx I \varepsilon c_B(x, t_0)$, where $\varepsilon$ is the molar absorption coefficient. According to the second assumption of the model, the evolution of the heat source distribution follows that of the mass distribution and asymptotically approaches an equilibrium distribution $h_{\mathrm{eq}}$.

The surface temperature can be factorized as $\Theta(0, \mathrm{t}) = \Theta_{\mathrm{eq}}(0) - \sum_{j=0}^{\infty} B_j \exp\left[-\frac{t_0}{\tau_j}\right] \Theta_j(0)$, which mathematically resembles the Crank solution for mass transport and predicts a quasi-exponential evolution of the FSD signal in the investigated system. The result establish a direct and linear relationship between the PA signal and the pigment concentration in the membrane in the optically transparent regime. More detailed information on the model can be found in the referenced paper.

### 2.4.2 DR technique

To relate the evolution of the PA diffusive-reflectance signal to the mass transport process, the effect of pigment accumulation at the membrane interface due to the molecular diffusion is considered as the first approximation.

We assume that the pristine membrane (i.e., prior to pigment penetration) is characterized by an initial reflectivity $R_0$, determined by its intrinsic scattering properties and refractive index mismatch. During the diffusion process, pigment molecules accumulate near the surface, introducing additional absorption while only weakly affecting scattering. As a consequence, photons contributing to the backscattered signal experience an increased probability of absorption before exiting the medium.

Under these assumptions, the effect of the pigment-enriched surface layer can be approximated as an effective absorbing filter. The diffusive reflectance signal can then be expressed as: $R(t_0) = R_0 \exp\left[-2\varepsilon \int_0^{l_{\mathrm{eff}}} c_b\,(x, t_0)\, dx\right]$, where $l_{\mathrm{eff}}$ denotes the effective photon penetration depth in the backscattering geometry. The factor of 2 accounts for the average photon path traversing the absorbing layer twice (inward and outward). In the case where the absorption is limited to the near-interface region ($x_{IF} = 0, l$ for acceptor/air and donor acceptor interfaces, respectively) and varies slowly over the scale of $l_{\mathrm{eff}}$, the above expression can be reduced to exponential formula $R(t_0) \approx R_0 \exp[-2\varepsilon l_{\mathrm{eff}}\, c_B(x_{IF}, t_0)]$. It should be emphasized that the pigment concentration at the interface, $c_B(x_{IF}, t)$, may deviate from the value predicted by the bulk diffusion model. In particular, the interfacial region can exhibit specific physicochemical interactions, such as adsorption, binding, or local structural heterogeneity, which are not accounted for within the standard diffusion framework.

In the diffusive DR configuration, the radiation backscattered from the sample is redirected into the photoacoustic chamber, where it is absorbed by a carbon black layer. The carbon black acts as the actual source of the photoacoustic signal. Within the RG photoacoustic framework [26], the generated signal is directly proportional to the modulated optical power absorbed by the transducer medium. In the present configuration, this corresponds to the intensity of the light returning from the sample surface. Consequently, the detected PA signal can be expressed as $S_{\mathrm{DR}} \propto I_{\mathrm{back}}$, where $I_{\mathrm{back}}$ is the intensity of the backscattered radiation incident on the carbon black layer.

Since the backscattered intensity is determined by the diffusive reflectance of the sample, the DR signal is directly proportional to the sample reflectance: $S_{\mathrm{DR}} \propto R(t_0)\, I$. Any changes in the optical properties of the sample, like increased absorption due to pigment accumulation, modify the backscattered intensity and the amplitude of the photoacoustic signal. Finally the configuration effectively converts variations in diffusive reflectance into a photoacoustic response, enabling indirect monitoring of optical absorption changes through the PA signal generated in the carbon black layer.

### 2.4.3 Transmission technique

The technique as considered in this work can be employed to stationary systems, as it measures the light transmissivity through the whole sample. The approach is very similar to the transmission measurements in the conventional spectrometers. In particular, the experimental arrangement consists of two sequential measurements. In the reference measurement, the modulated excitation light passes through the empty sample position and a photoacoustic chamber, where it is absorbed by a carbon black layer. As in the diffusive reflectance configuration, the carbon black acts as an efficient broadband absorber and serves as the source of the PA signal. The recorded signal therefore represents the spectral distribution of the incident radiation, $S_{\mathrm{ref}}(\Lambda) \propto I(\Lambda)$. In the actual measurement, the sample is introduced into the optical path. The incident radiation propagates through the sample and is attenuated according to the Lambert- Beer law. The transmitted light then enters the photoacoustic chamber and is absorbed by the carbon black layer, generating a PA signal proportional to the transmitted intensity: $S_{\mathrm{tr}}(\Lambda) \propto I_{\mathrm{tr}}(\Lambda)$. By taking the ratio of the transmitted and reference signals, the spectral transmittance of the sample is obtained: $T(\Lambda) = \frac{S_{\mathrm{tr}}(\Lambda)}{S_{\mathrm{ref}}(\Lambda)} = \frac{I_{\mathrm{tr}}(\Lambda)}{I_0(\Lambda)}$.

Assuming Beer–Lambert attenuation, the transmitted intensity can be expressed as $I_{\mathrm{tr}}(\Lambda) = I(\Lambda)\exp[-\alpha(\Lambda)l]$, and the measured transmittance: $T(\Lambda) = \exp[-\alpha(\Lambda)l]$.

This configuration effectively converts optical transmission into a photoacoustic signal via the carbon black absorber, enabling indirect measurement of spectral attenuation. In contrast to the FSD and DR modes, the transmission geometry probes the entire sample thickness, making it sensitive to the depth-integrated absorption rather than predominantly near-surface effects.

## Materials and methods

The experiments performed involved photoacoustic detection of dithranol (in a pharmaceutical form) permeation through a composite dodecanol/collodion (DDC) membrane. Three types of experiments were performed: stationary PA spectroscopy in transmission mode, and wavelength-fixed time-resolved MBPA in standard and diffuse-reflection modes. These experiments were aimed to deliver information on the spectral properties and characterization of the system (transmission PAS), mass permeation dynamics (FSD MBPA) and interfacial mass behaviour (DRPA in the multibeam mode).

The drug-membrane system investigated has been already considered as a candidate for a photoacoustics *proof-of-concept* studies due to its relatively slow dynamics (the lock-in detection time << permeation dynamics characteristic time), single photodegradation path: dithranol $\overset{hv}{\rightarrow}$ danthrone, very high light-to-heat conversion efficiencies for the pigments ($\eta_{dith} \approx \eta_{dant} \approx 1$ for the 0.1 mM pigment/dodecanol solution at 25 ºC), and non-overlapping absorption bands in the observed 260 - 500 nm region for the membrane (< 280 nm), dithranol (first observable peak ~280 nm, second/characteristic ~360 nm) and danthrone (first ~280 nm, second/characteristic ~430 nm) allowing to track the permeation behaviour of the species separately. The optical and photoacoustic characterization of the system can be found in [27].

### 3.1 Equipment

The experimental rig consisted of commercially-available components. For the spectroscopy studies, the light source module consisted of 900 W Xe light source (Newport 66921) and grating monochromator with adjacent mechanical chopper (Newport 74125), PAC 300 PA cell and a single lock-in amplifier (Signal Recovery model 7265). For the MBPA and DR studies the probing light was generated by LED modules equipped with 280, 360, 420 and 530 nm Lumiled Luxeons (with FWHM less than 20 nm, and peak intensities up to 10 W $m^{-2}$ in the linear regime), and controlled by (FG) function generator (Siglent 2042X). The light intensities were monitored by a photodetector (Thorlabs, PDA10A2 and PM100D2). The light intensity variations were generally less than 1% / hour, which is considered as of negligible impact on the acquired data quality. The probe beams were routed to the PA cell via optical cage system with dichroic mirror; the output signal was split into two independent lock-in units (Stanford Research System SR850) and analysed with respect to the reference frequency provided by the function generator unit. The time constant for the lock-in process was set to 1 s, the time interval between the measurements was set to 10 s; the filter slope was set to 18 dB/oct, which gives the detection bandwidth of 93 mHz.. By referring to the performance and reliability tests of the measurement system presented in [20], the modulation frequency stability can be estimated as 0.4 Hz. The two yields the instrumental limit for the multibeam detection. A diagram for the MBPA setup is shown in Fig. 1.

Before the actual experiment, the total instrumental noise of the measurement system was determined. For this purpose, the signal was measured using exactly the same hardware configuration as in the experiment, with the only difference being that the inlet to the photoacoustic chamber was blocked with a thick blackened plate. The noise magnitude was in order of microvolts, i.e. approximately 4 μV at 35 Hz and 1 μV at 120 Hz.

The driving voltage ranges for the LEDs were adjusted to maintain linear light intensity response of the units, a condition necessary to keep the PA responses independent. For the MBPA sorption studies, an additional condition for equal light fluxes (carbon black response to 360 and 430 nm beams adjusted by the $h\upsilon_{360\,nm}/h\upsilon_{430\,nm}$ ratio, under assumptions of $\alpha_{CB}(360\,nm) = \alpha_{CB}(430\,nm)$ and $\eta_{CB} = 1$) was applied. By keeping in mind similar and very high light-to-heat conversion efficiencies for dithranol and danthrone, applying beams of similar fluxes allows for direct comparison of molecular contents.

The photoacoustic non-resonant cell PAC 300 with appropriate heads for PAS, DR and transmission modes was delivered by MTEC Photoacoustics. The cell was equipped with B&K 4176 microphone. The preamp gain was set to 20x. According to the producers data sheets, dynamic range for the setup covers the range of ~ $10^{-5}$ V – 1 V. For the DR head, the sample radius was 6 mm, and the sample – carbon black o-ring (of thickness of 1 mm) distance was 2 mm. As such, the limiting angle of incidence for the light probing rough surface was not larger than ~ 35°. Although for the (0°; 35°) interval the s- and p-type reflectivities change dramatically (assuming $N_2 = 1.5$: $R_s(0^o) = 0.04$ and $R_s(35^o) = 0.066$, $R_p(0^o) = 0.04$ and $R_p(35^o) = 0.02$), their average remains relatively stable ($R_{eff}(0^o) = 0.04$ and $R_{eff}(35^o) = 0.042$). This implies almost homogeneous response from the sample surface in case of non-polarized probe beams.

Data for instrumental hum and exemplary Lumiled light intensity distribution and variations can be found in the Supplementary Materials.

### 3.2 Sample preparation and the experimental procedures

2% dithranol/Vaseline suspension was delivered by a local pharmacy. Pure dithranol, 1-dodecanol and 4-8% collodion solution were purchased from Sigma-Aldrich (PNs: D2773, 443816, 09986).

#### 3.2.1 Membrane preparation

Dodecanol–collodion (DDC) membranes were synthesized by mixing 25 g of the collodion solution with 21.5 mL of an EtO/EtOH solvent mixture and 0.53 g of 1-dodecanol. The final dry mass ratio of dodecanol to nitrocellulose was approximately 1:4. The resulting solution was cast onto a leveled glass plate (3 cm × 5 cm × 0.2 cm). After 30 min of solvent evaporation under ambient conditions, the film was transferred to a desiccator and dried for 24 h.

After drying, the membrane was gently removed from the plate. Its thickness was measured at five randomly selected locations using a Mitutoyo 2110S-10 dial indicator to verify uniformity. The final membrane thickness was $l_{\mathrm{DDC}} = 15 \pm 1\,\mu$m.

Circular membrane samples (discs) of 0.8 cm and 0.6 cm in diameter were cut from the same membrane sheet using a sharp-edge tool. The larger discs were used for front-surface detection (FSD) and transmission measurements, while the smaller ones were dedicated to diffusive reflectance (DR) experiments.

### 3.2.2 Transmission measurements – stationary systems

Three types of samples were investigated: (i) pristine membranes, (ii) pigment-loaded membranes, and (iii) membranes containing photodegraded pigment.

For pigment loading, membrane samples were immersed in a 1 mM dithranol solution for approximately 24 h, which was assumed sufficient to reach saturation. For photodegradation studies, the saturated membranes were removed from the solution and exposed to 360 nm radiation for 2.5 h. The irradiation intensity and modulation frequency were adjusted to match the conditions used in the sorption experiments.

Transmission measurements were performed using a dedicated transmission PA head supplied by the instrument manufacturer (Fig. 1c). The membrane samples were placed on a thin quartz plate and positioned above the photoacoustic chamber, thereby protecting the KBr window sealing the chamber.

### 3.2.3 Front-surface detection (FSD) – dynamic system

FSD measurements (Fig. 1a) were carried out using a standard PA head provided by the manufacturer. A custom-made quartz cuvette (diameter: 0.8 cm, height: 0.4 mm), matched to the dimensions of the PA head, was filled with the donor medium (2% dithranol/Vaseline suspension) and allowed to stabilize for 30 min inside the head. Subsequently, the membrane (acting as the acceptor) was placed onto the donor surface, and the assembly was immediately transferred into the photoacoustic chamber to initiate the measurement.

### 3.2.4 Diffusive reflectance (DR) – dynamic system

DR measurements were performed using the dedicated reflectance head supplied by the manufacturer. In this configuration, the donor suspension was placed in a metallic sample cup provided with the system. After a 30 min stabilization period, the membrane was placed onto the donor medium, and the head was promptly positioned in the PA chamber to start data acquisition.

### 3.2.5 Measurement protocol

All experiments were performed using membranes originating from the same synthesis batch to ensure consistency. In all time-dependent experiments, the delay between placing the membrane on the donor medium and the start of signal acquisition was kept below 20 s. After initiating the measurement, the operator left the room to minimize external disturbances. Data were monitored and preliminarily analysed in real time using remote access. Each experiment was continued until a steady-state signal was reached, but not longer than 90 min.

The equipment warm-up time was not less than 30 min. Prior to each measurement series, system stability was verified by recording repeated reference spectra. Transmission spectra were acquired at a modulation frequency of 120 Hz over the spectral range of 260–700 nm. A high-pass filter was applied for wavelengths above 500 nm.

Time-dependent measurements were performed using quasi-monochromatic LED sources driven by a dual function generator, with constant average light intensity maintained throughout the experiment. For the FSD studies, modulation frequencies in the range of 20–900 Hz were tested; however, only the ~50 Hz component (corresponding to the membrane thickness) was selected for further analysis. In the case of DR measurements, two distinct modulation frequency ranges were employed to minimize potential environmental influences. The signal from the PA cell was routed into two lock-in units working in parallel, with proper reference frequencies provided by the FG unit. The signal vs time was recorder by LabView-based software.

In each experiment, the probe light beam spot was adjusted to cover the whole sample area. As such the PA signal is considered to homogenize the sample response.

For the MB FSD experiments, the separation frequency was set above the instrumental 0.4 Hz limit to minimize the possibility of interference effects. The goal was, however, to preserve the sampling depth accuracy above the 95% threshold.

The measurements were performed at the room temperature of 25 ± 1 ºC maintained using an air-conditioning system.

### 3.3 Temperature effects

The impact of sample heating due to probe light absorption on the transport system characteristics (e.g., diffusion coefficient and decay rates) can be estimated as follows. We assume that the membrane and the donor suspension are in perfect thermal contact, such that the effective thermal response of the irradiated region is dominated by the properties of the suspension (approximated as petroleum jelly). Taking an average irradiance of $5\ \mathrm{W\ m^{-2}}$, an irradiated area on the order of $10^{-5}\ \mathrm{m}^2$, and a sample volume comparable to that of the quartz cuvette, together with the density and specific heat capacity of Vaseline, the resulting temperature increase is estimated to be on the order of $10^{-3}\ \mathrm{K\ s^{-1}}$.

This estimate represents an upper bound, as it neglects heat dissipation through thermal contact with the quartz cuvette and the metallic components of the photoacoustic system base plate. Inclusion of these effects would further reduce the temperature rise. Therefore, thermal effects induced by the probing light can be safely neglected in the analysis of transport properties.

## 4 Results and discussion

### 4.1 Preliminaries

The preliminary studies involved spectral characterization of the membrane (test 1), the pigment/membrane (test 2) and the photodegraded pigment/membrane (test 3) samples in the PA transmission mode, as schematically shown in Fig. 1c. The experimental findings are shown in Fig. 2. It can be noted that the DDC membranes (blue triangles) absorb light in the UV region, below 330 nm, and exhibit ~ 30% transmittance for 280 nm in the case of the 15 μm thick samples. This implies a characteristic absorption half-value layer (a layer which reduces the light intensity by 50%) HVL of ~ 8.5 μm or equivalently the light absorption coefficient of $\alpha_{test\ 1}(280\ nm)$ = 0.08 μm$^{-1}$.

The dithranol/DDC sample (black squares) exhibits a characteristic peak at 360 nm, in line with our previous studies with the use of standard front-side detection PAS technique [27]. The transmittance value obtained for the sample implies HVL of ~20 μm, $\alpha_{test\ 2}(360\ nm)$ = 0.03 μm$^{-1}$ and characteristic absorption length $1/\alpha_{test\ 2}(360\ nm)$ of around 33 μm, which is greater than the sample thickness. The latter fulfils the condition for the PA depth-profiling studies, as discussed in Section 2.2.

The photodegraded sample exhibits broad absorption region between 360 and 460 nm. This can be attributed to two pools of dithranol photoproducts: the membrane matrix-bound pool of dithranol molecules with absorption band shifted to ~ 400 nm, and photodegradation product – danthrone (with the absorption peak ~ 430 nm) molecules. Both types of the photoproducts were already recognized during the drug desorption studies reported in [27]. Taking the ~ 70% transmission magnitude at 430 nm site due to the danthrone presence, then: HVL is 29 μm, $\alpha_{test\ 3}$ (430 nm) = 0.024 μm$^{-1}$ and finally $1/\alpha_{test\ 3}(430\ nm)$ is 42 μm.

It is worth to mention, that for both cases of test 2 and test 3 samples, the presence of permeant leads to a certain drop in the light transmission below 330 nm. For all of the samples considered, for the spectral region of 500-600 nm the transmission magnitude was above to 95%, without a presence of any characteristic absorption bands (data not shown).

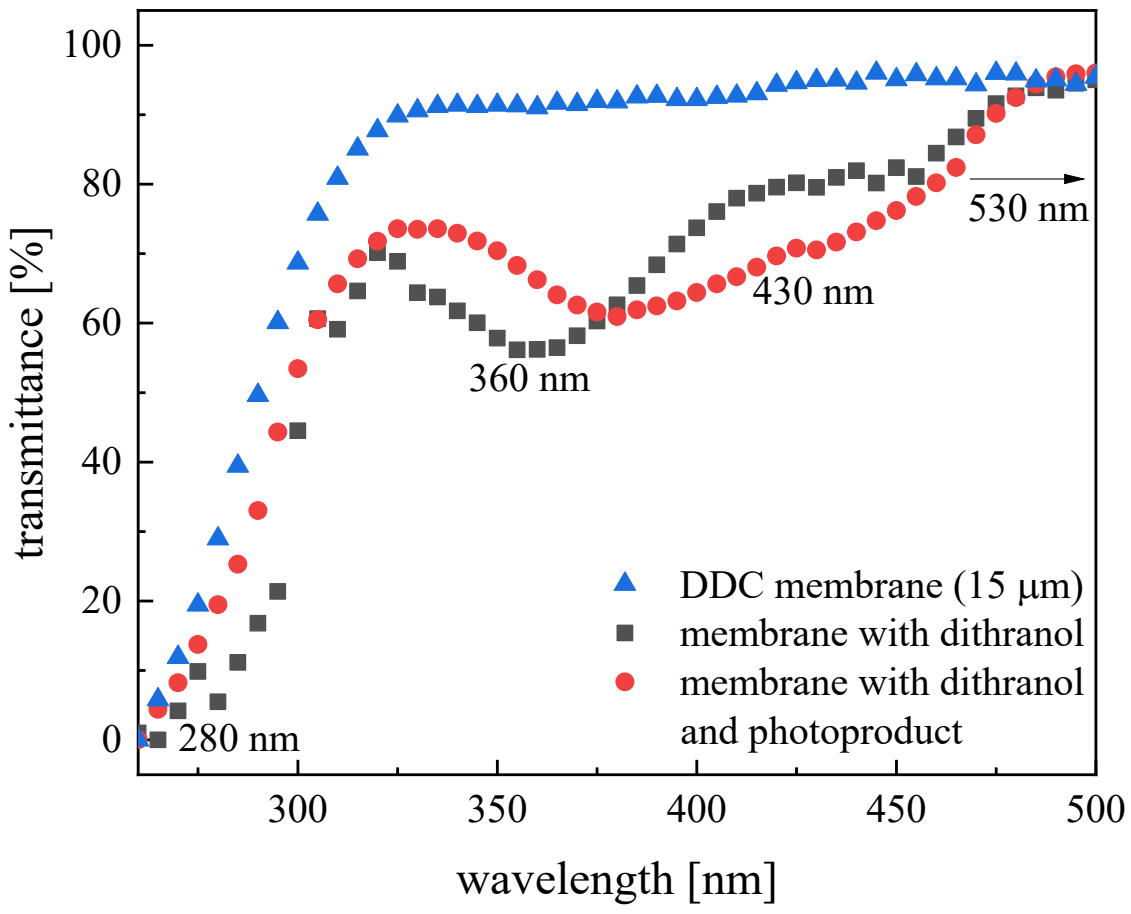

Fig.2 Stationary PA transmission spectra for a pure acceptor membrane (blue triangles, 15 μm thick), the membrane soaked with dithranol molecules (black squares) and the membrane soaked with the pigment and exposed to 360 nm irradiation for 2.5 hr, exhibiting the presence of a photoproduct (red dots). The spectral properties of the system underline the characteristics of the light sources desired for the transport experiments.

The preliminary results obtained can be summarized by relating specific spectral regions to the type of information obtainable about transport processes via the introduced PA modalities. This includes:

a) for $\Lambda$ < 300 nm low transmission values are observed, with the HVLs < $l_{DDC}$, as such, the region is not appropriate for the sorption kinetics studies via depth profiling. In the case of DRPA, the amount of light backscattered from the air/membrane interface dominates that from the membrane/donor interface. As such this region appears to be suitable for investigating the upper interface sorption kinetics.
b) Moderate transmission values are observed for the saturated/degraded samples at 330 nm < $\Lambda$ < 450 nm (i.e. the region with the characteristic absorption bands of the permeants), with 1/α(Λ)) > $l_{DDC}$. This indicates a spectral region that is suitable for the PA depth profiling and investigating the *volumetric* sorption kinetics.

c) Above 450 nm (to ~ 600 nm) the light transmission is very high and lacks of any characteristic absorption patterns, with the light intensity loss probably via light scattering or residual (negligible) absorption. As the light penetrates through the whole membrane, the backscattered light brings information on both of the air/membrane and the membrane/donor interfaces. Finally, the region appears to be suitable for the DRPA studies.

The findings justify the selection of the light sources for the depth profiling (360 and 430 nm) and DRPA (280 and 530 nm) studies.

## 4.2 MBPA-based sorption profiles and the diffusion coefficient

### 4.2.1 Basic approach to the mass release kinetics studied with MBPA

A single beam PA modality with the front side detection has been employed into release kinetics studies for several times, with different irradiation sources ranging from UV to IR [17,28,29]. In a typical experiment, an acceptor (membrane) is placed on a pigment donor substrate; a concentration difference between the subsystems triggers molecular diffusion often considered as of Fickian character. By benefiting from the depth profiling PA method, whereby the screening depth is adjusted by setting the correct light modulation frequency, it is possible to monitor the total pigment content within a profile of interest. Figure 3a shows a general diagram of the mass release system used in photoacoustic studies.

The MBPA mass release experiment, which examined pigment transfer from a donor substrate into an acceptor membrane using 360- and 430-nm probe beams, was performed using the configuration shown in Fig. 3a. Modulation sweeps were performed between 30 logarithmically spaced frequencies ranging from 20 to 900 Hz. The modulation frequencies matching the whole 15 μm membrane profile were found to be ~50 Hz, i.e. a cut-off frequency where the initial scan exhibited pseudo-linear character in the log/log (signal vs frequency) representation. These values allow the thermal diffusivity of the material to be determined, $\beta_s \approx 3.5 \cdot 10^{-8}$ m$^2$ s$^{-1}$, in line with our previous studies using similar procedures and materials [27]. The separation frequency between the beams was set to a safe level of $\Delta f = 2$ Hz, offering 98% sampling depth accuracy for frequencies greater than ~49 Hz. The experimental results for the membrane sorption experiment are shown in in Fig. 3b. Both data sets (360 and 430 nm) exhibit pseudo-exponential character, in line with the prediction of the EC-based diffusion model. In fact, many analytical schemes for the membrane transport quantification refer the specific solutions to the diffusion equation under various initial and boundary conditions as provided in the Crank's textbook [25], neglecting at the same time reaction processes. The list includes closed-form solutions, but also approximations valid for short times like square-root and exponential functions.

By fitting EC-based exponential model to the experimental data two characteristic times for the 360 nm and 430 nm sets can be determined: $\tau^{360} \approx 816$ s and $\tau^{430} \approx 2241$ s, which for the membrane of thickness of 15 μm translate to the diffusion coefficients of ~1.05 · $10^{-9}$ cm$^2$ s$^{-1}$ and ~3.8 · $10^{-10}$ cm$^2$ s$^{-1}$. The basic approach raises questions about the origin of the discrepancies in the diffusion coefficients between structurally comparable molecules, and the impact of photodegradation rates and the degradation location on the reliability of the determined transport parameters. This motivates the incorporation of MBPA combined with diffusion-reaction modelling to determine transport parameters in membrane permeation experiments.

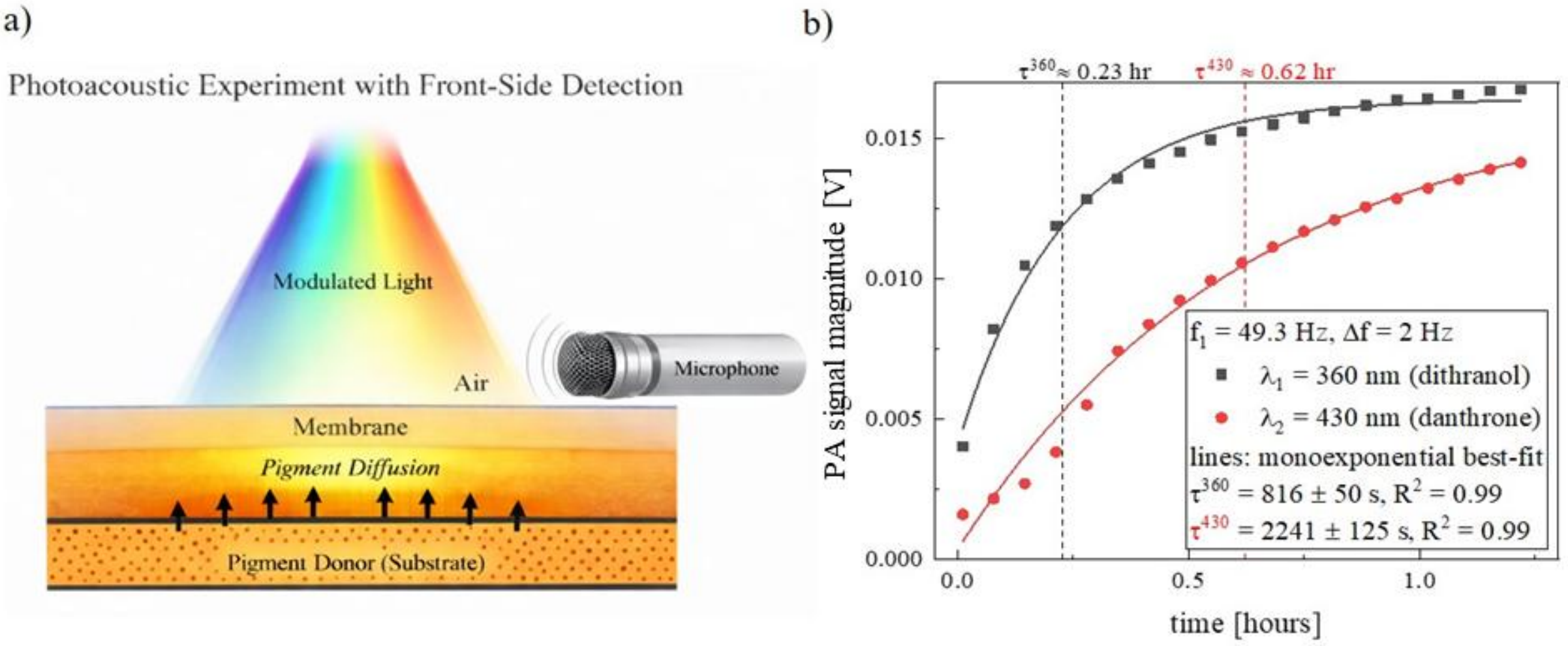

Fig.3 a) an overview on the plane-sheet donor → acceptor mass transport problem symmetry for the front side detection scheme; b) MBPA release kinetics data acquired for: a 360 nm source at 49.3 Hz and a 430 nm source at 51.3 Hz. Both data sets revealed an exponential character, which allowed the characteristic time scales of the underlying mass transport processes to be determined.

#### 4.2.2 EC-based model for a multispecies mass transport with decay

We consider a non-equilibrium system consisting of two compartments $A$ (of reduced volume $V_A \equiv l_A$) and $B$ ($V_B \equiv l_B$) in contact and undergoing mass exchange of trace molecules $n_1$. The trace molecules concentration ($c = n/V$) difference between the compartments leads to the molecular diffusion of magnitude given by the EC-based parameters $\frac{c_{B1}(t)-c_{A1}(t)}{R} = \Sigma_B n_{B1} - \Sigma_A n_{A1}$, where $R$ stands for the $B$ system resistivity ($R = \frac{l_B}{D^{EC}}, \Sigma = 1/Rl$). Beside the diffusion, we assume the molecules can undergo a degradation processes characterized by the time-independent decay constants λ, specific for each compartment. The degradation leads to the appearance of a second pool of molecules in both subsystems ($n_{A2}$ and $n_{B2}$), which can also trigger molecular diffusion in the presence of a proper concentration gradient. As such it is assumed, that the mass transport kinetics of the system can be described by four parameters coupling the molecular populations in both compartments, as schematically drawn in Fig. 4.

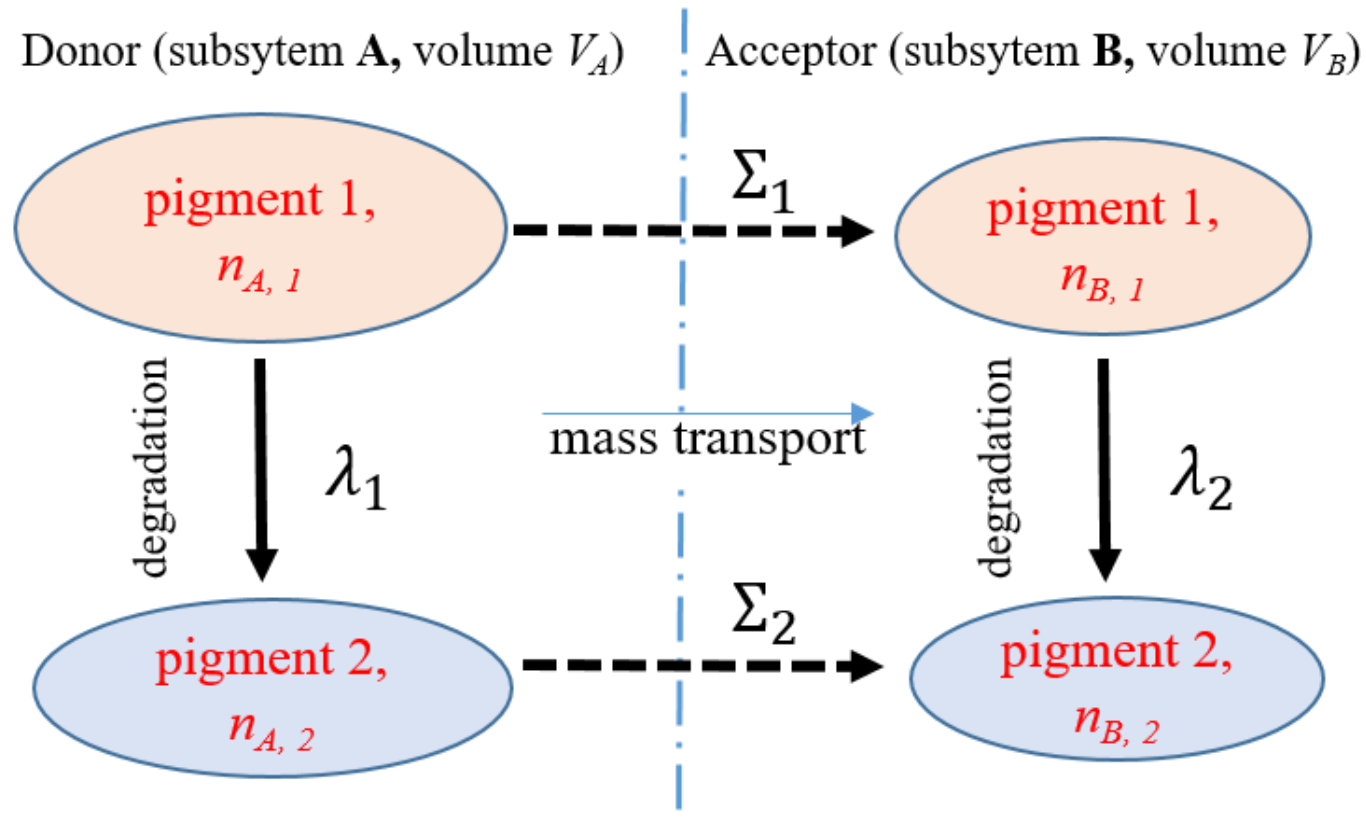

Fig. 4 Block diagram for an EC-based diffusion-reaction model, involving two subsystems (donor $A$ and acceptor $B$) and two pigments pools, yielding 2x2 populations coupled by intersystem transfer ($\Sigma_{1,2} = \Sigma_B n_{B1,2} - \Sigma_A n_{A1,2}$) and decay rates ($\lambda_{1,2}$).

By referring to Eq. 2, the system evolution can be specified by:

$$\frac{d}{dt}\begin{pmatrix} n_{A1} \\ n_{A2} \\ n_{B1} \\ n_{B2} \end{pmatrix} = \begin{pmatrix} -(\lambda_1 + \Sigma_{A1}) & 0 & \Sigma_{B1} & 0 \\ \lambda_1 & -\Sigma_{A2} & 0 & \Sigma_{B2} \\ \Sigma_{A1} & 0 & -(\lambda_2 + \Sigma_{B1}) & 0 \\ 0 & \Sigma_{A2} & \lambda_2 & -\Sigma_{B2} \end{pmatrix} \begin{pmatrix} n_{A1} \\ n_{A2} \\ n_{B1} \\ n_{B2} \end{pmatrix} \quad (9)$$

or equivalently $\frac{d\boldsymbol{n}}{dt} = M\boldsymbol{n}$, where $\boldsymbol{n}$ is the population vector and $M$ stands for the coefficient matrix composed of constants $\Sigma$ and $\lambda$. The equation system is linear and can be solved explicitly when the initial conditions are specified, i.e. $\boldsymbol{n}(t) = \exp[Mt]\boldsymbol{n}(0)$, where $\boldsymbol{n}(0)$ stands for the vector of initial conditions. The solution of the linear system can be expressed as a superposition of eigenmodes: $\boldsymbol{n}(t) = \sum_i C_i \boldsymbol{v}_i \exp[\chi_i t]$, where $\chi_i$ and $\boldsymbol{v}_i$ are the eigenvalues and corresponding eigenvectors of the system matrix $M$, while $C_i$ are fixed by the initial condition of the system. Some exemplary cases for *c(t)* sorption kinetics are presented in Fig. 5. In particular, Fig. 5a demonstrates the mass transport of chemically-stable two types of trace molecules (i.e. decay constants $\lambda_1 = \lambda_2 = 0$) between two volumetrically similar subsystems (of the reduced volume of $l$ = 1 μm) obeying initial conditions: $c_{A1}(0) = c_{A2}(0) = 2$ and $c_{B1}(0) = c_{B2}(0) = 0$. Two distinct resistivities for the mass transport are considered, i.e. $R_1 = 2R_2 = 0.5$ hr μm$^{-1}$. As expected the equilibrium concentrations for both species in subsystems $A$ and $B$ are equal (to be more precise, for a system studied the mass spreads equally between the subsystems in the $t \to \infty$ limit, and so $c_{A1}(\infty) = c_{B1}(\infty)$ and $c_{A2}(\infty) = c_{B2}(\infty)$, however the transport kinetics strongly depends on the resistivities; by recalling the definition of the characteristic time $\tau$ ($= Rl_{eq}$, where $l_{eq}$ stands for the reduced equivalent volume, here 0.5 μm), that is a time for the system to reach its ~ 63 % of equilibrium value, $\tau_1 = 0.25$ hr and $\tau_2 = 0.125$ hr. Fig. 5b demonstrates case with the initial conditions as before, but with decaying and *mobile* 1st pool of trace molecules, and stable ($\lambda_2 = 0$) and relatively immobile 2nd pool of molecules (in the time window considered as $R_2 \gg R_1$). The generation and presence of the second pool of molecules is noticed mainly in the $A$ compartment, where due to $\lambda_1 > 0$ the magnitude of $c_{B1}$ is hindered with respect to the Fig. 5a case. It can be noted, that for long times one observes the $c_{B1}$ excess over $c_{A1}$, which eventually leads to the back-diffusion process ($B \to A$). Fig. 5c shows two pools of molecules that initially occupy a large donor compartment ($c_{A1}(0) =$ $2c_{A2}(0) = 2$) and exchange (for $t > 0$) with a small acceptor compartment at a similar rates ($R_1 = R_2$), and with the decay only in the $B$ compartment. Here one recognizes rapid absorption kinetics in the $B$, and slow release in the $c_{A1}$ behaviour. Due to the $\lambda_2$ decay mode, the concentration drop can be also seen in the $c_{B1}$ kinetics for $t > 0.4$ hrs. On contrary, the $c_{A2}$ for $t > 0.15$ hrs (interception of $c_{A2}$ and $c_{B2}$) appears to rise due to the back-diffusion. Fig. 5d presents a situation where $c_{A1}(0) = 1$ and $c_{A2}(0) = c_{B1}(0) = c_{B2}(0) = 0$, with $l_A \gg l_B$, $R_1 = R_2$ and $\lambda_1, \lambda_2 > 0$. This conditions, at least to some extent, match conditions for photodegradable pigment transfer (pigment and its photoproduct remain similar chemically and structurally, which indicates similar resistivities) from large reservoir to a thin membrane (chemical difference between the compartments affects $\lambda$), as experimentally studied earlier. By selecting proper values for the transport parameters, including $R_1 = 0.02$ hr μm$^{-1}$ (which, along with the membrane thickness of 15 μm, yields $\tau = R_1 l_B = 0.3\ hr = 1020$ s, is close to the τ[360] reported earlier) and $\lambda_2 \gg \lambda_1$ (degradation mainly in the acceptor) it is possible to capture crucial properties for the observables $c_{B1}$ $c_{B2}$ in Fig. 3. This includes concentration quasi-equilibration in $B$ for the 1st molecules pool (dithranol, black points in Fig. 3 and green line in Fig. 5d), and two inflection points for the 2nd molecules pool (danthrone, red points, red line).

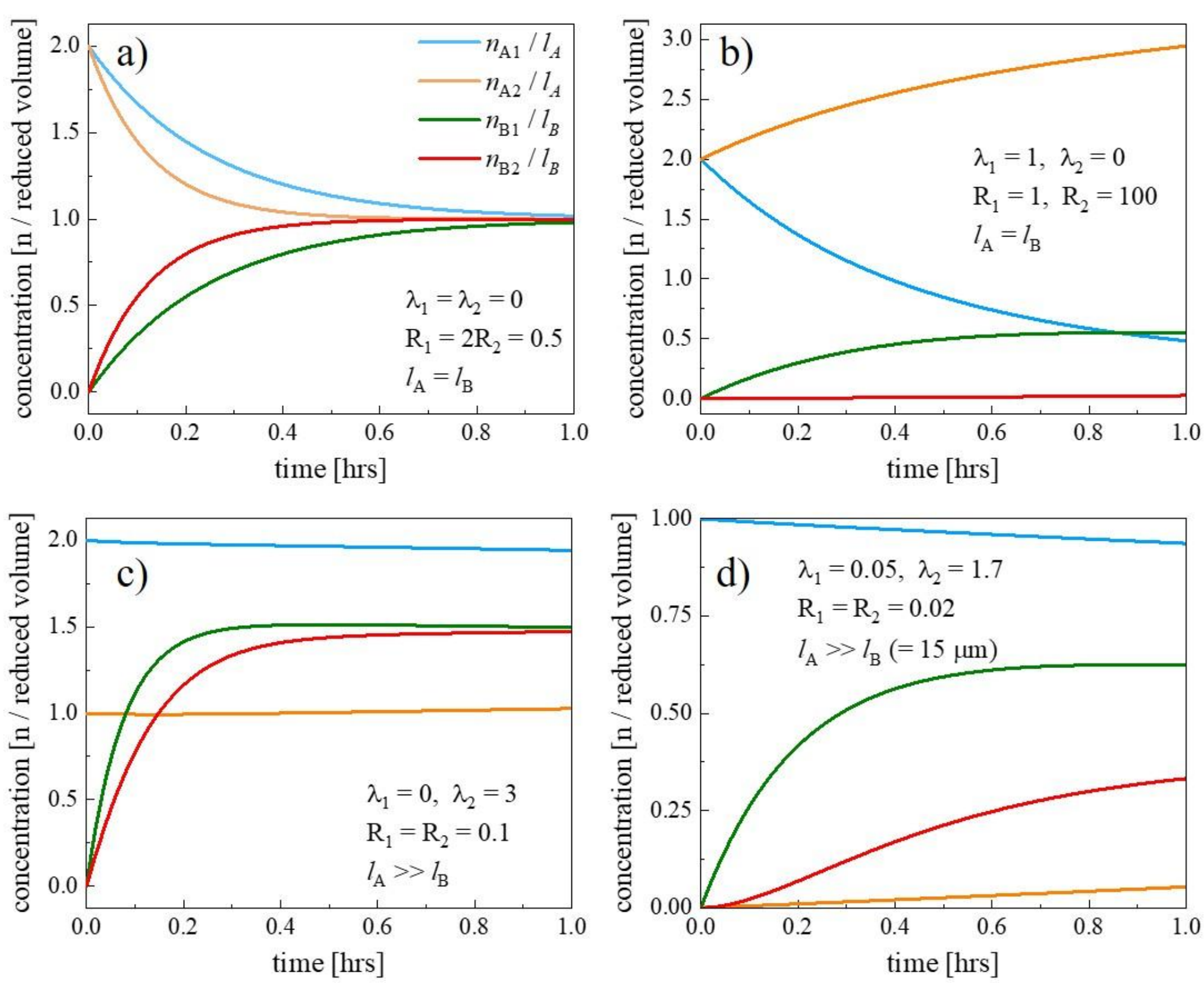

Fig. 5 Results of the EC diffusion-reaction model simulations for the pigments concentration evolutions in the donor and acceptor phases: a) independent diffusion of the pigments characterized by distinct but of the same order diffusion coefficients; b) decay and pseudo-binding ($R_2 \gg R_1$) in the donor phase; c) fast transport accompanied with the decay in the thin acceptor phase, signatures of the back diffusion can be noticed in the $c_{B1}$ behaviour; d) a model for a membrane transport as studied during PA experiments: similar transport rates for the two specimens, decay mainly in the acceptor phase. Case a) and b) - compartments volumetrically equal, c) and d) – thick donor in contact with a thin acceptor. λ*s* and *Rs* are given in hr and hr μm$^{-1}$, respectively.

The main dissonance between the experimental data and simulations involves relative magnitudes of the PA signals directly proportional to the concentrations and the concentrations itself. As discussed in the Theory section, the PA signal magnitude is proportional to the product of light intensity, pigment concentration and the light-matter interaction cross-section via the light absorption coefficient, and the thermal deactivation efficiency coefficient. In the experiment, the light intensities (360 and 430 nm) were controlled to provide similar photon fluxes on average; the light absorption coefficient in the quasi-equilibrium states were estimated to $\alpha_{test\ 2}(360\ nm)$ = 0.03 μm$^{-1}$ and $\alpha_{test\ 3}$ (430 nm) = 0.024 μm$^{-1}$, but appropriate corrections for the absorption coefficients do not solve the issue. At this point it can be mentioned, that photoacoustic spectra for 1 mM pure dithranol and danthrone solutions in dodecanol (normalized with respect to the reference CB sample) presented in [27] also showed distinct PA magnitudes at the characteristic absorption wavelengths, with the effect attributed to different $\eta$ or/and the light-matter interaction cross-sections, in line with the current conclusions. Moreover it is known that in the presence of light dithranol molecules bind to the DDC membrane matrix, which spectral fingerprint can be found around 400 nm. The effect has been studied in [27] and recognized here (PA transmission spectra, Fig. 2). This implies that in order to describe the process, the model should be further extended by introducing new pool of immobile molecules (large/infinite resistivity) $n_{B3}$ as an additional photoproduct of the first pool of pigment molecules in the $B$ compartment, generated at the rate of $\lambda_3$. Although it is technically possible to introduce a number of new pools and light sources to investigate the populations experimentally, this is limited by the sources and molecular bands full width at half maximum parameter (FWHM) and the Rayleigh criterion rule, and yields difficulties when applied to a quite narrow spectral range of interest.

We now focus on the short-time regime, in which the transport kinetics can be quantified from the initial slopes and curvatures of the release profiles. The goal is to obtain closed-form approximations that enable direct estimation of the governing transport parameters. In this regime, the amount of tracer accumulated in the acceptor compartment is assumed to remain small compared to that in the donor. Consequently, the driving concentration difference can be approximated as $c_A - c_B \approx c_A$, which significantly simplifies the analytical treatment. To underline an approximate character of the coefficient matrix, we interchange $\Sigma_{A1}$ and $\Sigma_{A2}$ with $\sigma_1$ and $\sigma_2$ respectively, and set $\Sigma_{B1} = \Sigma_{B2} = 0$. Then the system evolution is given by:

$$\frac{d}{dt}\begin{pmatrix} n_{A1} \\ n_{A2} \\ n_{B1} \\ n_{B2} \end{pmatrix} = \begin{pmatrix} -(\lambda_1+\sigma_1) & 0 & 0 & 0 \\ \lambda_1 & -\sigma_2 & 0 & 0 \\ \sigma_1 & 0 & -\lambda_2 & 0 \\ 0 & \sigma_2 & \lambda_2 & 0 \end{pmatrix} \begin{pmatrix} n_{A1} \\ n_{A2} \\ n_{B1} \\ n_{B2} \end{pmatrix} \tag{10}$$

The solution for the $n_{A1}(t)$ is given by a single exponential formula:

$$\frac{dn_{A1}}{dt} = -(\lambda_1+\sigma_1)n_{A1} \rightarrow n_{A1}(t) = n_{A1}(0)\exp[-(\lambda_1+\sigma_1)t]. \tag{11}$$

For the $n_{12}(t)$ we start with the equation:

$$\frac{dn_{A2}}{dt} + \sigma_2 n_{A2} = \lambda_1 n_{A1}(t) = \lambda_1 n_{A1}(0)\exp[-(\lambda_1+\sigma_1)t], \tag{12}$$

which can be further rewritten and expanded with the use of the integrating factor into:

$$n_{A2}(t) = n_{A2}(0)\exp[-\sigma_2 t] + \lambda_1 n_{A1}(0)\int_0^t \exp[-\sigma_2 t]\exp[-(\lambda_1+\sigma_1)t]\,dt. \quad (13)$$

The solution for the above mentioned is given by:

$$n_{A2}(t) = n_{A2}(0)\exp[-\sigma_2 t] + \frac{\lambda_1 n_{A1}(0)}{\sigma_2-(\lambda_1+\sigma_1)}(\exp[-(\lambda_1+\sigma_1)t] - \exp[-\sigma_2 t]). \quad (14)$$

Similar steps can be taken to obtain solution for the $n_{B1}(t)$, which yields:

$$n_{B1}(t) = n_{B1}(0)\exp[-\lambda_2 t] + \frac{\sigma_1 n_{A1}(0)}{\lambda_2-(\lambda_1+\sigma_1)}(\exp[-(\lambda_1+\sigma_1)t] - \exp[-\lambda_2 t]). \quad (15)$$

Finally, the last population is given by a sum of $dn_{B2}/dt = \sigma_2 n_{A2}(t) + \lambda_2 n_{B1}(t)$, thus:

$$n_{B2}(t) = n_{B2}(0) + \sigma_2\int_0^t n_{A2}(t)dt + \lambda_2\int_0^t n_{B1}(t)dt. \quad (16)$$

By setting the following: $A_{A2} = \frac{\lambda_1 n_{A1}(0)}{\sigma_2-(\lambda_1+\sigma_1)}$, $B_{A2} = n_{A2}(0) - A_{A2}$, $A_{B1} = \frac{\sigma_1 n_{A1}(0)}{\lambda_2-(\lambda_1+\sigma_1)}$ and $B_{B1} = n_{B1}(0) - A_{B1}$, the evolutions of the populations can be simplified to:

$$n_{A2}(t) = A_{A2}\exp[-(\lambda_1+\sigma_1)\,t] + B_{A2}\exp[-\sigma_2 t]; \quad (17)$$

$$n_{B1}(t) = A_{B1}\exp[-(\lambda_1+\sigma_1)\,t] + B_{B1}\exp[-\lambda_2 t]; \quad (18)$$

$$n_{B2}(t) = n_{B2}(0) + B_{A2}(1-\exp[-\sigma_2 t]) + B_{B1}(1-\exp[-\lambda_2 t]) + \frac{\sigma_2 A_{A2}+\lambda_2 A_{B1}}{\lambda_1+\sigma_1}(1 - \exp[-(\lambda_1+\sigma_1)t]). \quad (19)$$

By considering the following: 1) the pigment and its photoproduct are structurally comparable (i.e. in terms of molecular radius affecting the diffusion coefficient via Einstein-Stokes relationship), their transport rate parameters are approximately equal, and so $\sigma_1 = \sigma_2$; 2) in many practical cases the initial condition for the system can be given by: $n_{A2}(0) = n_{B1}(0) = n_{B2}(0) = 0$, while $n_{A1}(0) > 0$; 3) by considering the system evolutions for short times, the exponents can be approximated using the first-order Taylor expansion, i.e. $\exp[-\lambda t] \approx 1 - \lambda t$. Under all of these conditions the observables $n_{B1}(t)$ and $n_{B2}(t)$ can be expressed by simple formulas:

$$n_{B1}(t) = \sigma_1 n_{A1}(0)t \quad \text{or} \quad c_{B1}(t)/c_{A1}(0) = G_F D^{EC} t \quad (20)$$

$$n_{B2}(t) = \sigma_1 n_{A1}(0)\frac{(\lambda_1+\lambda_2)}{2}t^2 \quad \text{or} \quad c_{B2}(t)/c_{A1}(0) = G_F D^{EC}\bar{\lambda}t^2 \quad (21)$$

where $\bar{\lambda}$ stands for the averaged decay rate parameter, $G_F$ is related to the system geometry ($= l_A/l_B^2 l_{eq}$), while the concentrations are given by $c = n/l$ ($l$ – the reduced volume). It can be noted, that for sufficiently thin acceptor compared to the donor phase the sigma becomes $\sigma_1 \approx D^{EC} l_B^{-2}$. The model needs extension for the apparatus constant when applied into raw experimental data analyses.

From Eqs 20-21 it is clear, that quantification of the transport parameters requires information on the initial pigment concentration in the donor compartment. In the case of a donor that is much larger than the acceptor volume, the initial concentration in *A* can be approximated with $c_A(0) = c_B(\infty)$ (i.e. by measuring the equilibrium concentration in the acceptor) as long as λ = 0. The latter can be achieved experimentally by measurement under weak light irradiation conditions. In fact, such experiments for similar systems as studied here were performed earlier, yielding diffusion coefficient of $D_{33\%} \approx 1.4 \cdot 10^{-9}$ cm$^2$ s$^{-1}$ for membrane of porosity ~ 33% approximated from the final dry mass ratios (for the membrane studied here the porosity is around 20%) [11]. By adopting the homogenization rules for the EC model [21] and applying correction for the porosity impact, $D_{20\%} \approx 0.85 \cdot 10^{-9}$ cm$^2$ s$^{-1}$, $D_{20\%}^{EC} \approx 2.22 \cdot 10^{-9}$ cm$^2$ s$^{-1}$ which corresponds to resistivity of $R_{20\%} = 1.87 \cdot 10^{-2}$ hr μm$^{-1}$ and finally $\tau \approx 0.28$ hr ≈ 1000 s, reasonably close to the values reported in Figs 3 and 5d.

By equating the linear best-fit slope with the experimental data acquired for short times (here: first 3-4 data points from the Fig. 3b, λ = 360 nm) it is possible to find the numerical value for the product of apparatus constant $C_{app}$ and the $G_F$ factor as predicted from the Eq. 20. For example, for the slope of 3.8·10$^{-2}$ mV hr$^{-1}$ (from the linear fitting to the first 3 experimental points) and $D^{EC} = D_{20\%}^{EC}$ of $2.22 \cdot 10^{-9}$ cm$^2$ s$^{-1}$, $C_{app}G_F = slope/D^{EC} \approx 4.7 \cdot 10^3$ mV cm$^{-2}$. The results can be further employed to evaluate the $\bar{\lambda}$ if the quadriatic prefactor of Eq. 21 is estimated by fitting the parabolic model to the experimental data (Fig.3b, λ = 430 nm). In particular, from the fitting of first 5 experimental points (where the parabolic character is revealed) to the $signal(t) = c_p t^2 + constant$ function, where $c_p = C_{app}G_F D^{EC}\bar{\lambda}$, the prefactor $c_p \approx 4.83 \cdot 10^{-2}$ mV hr$^{-2}$, and finally $\bar{\lambda}$ is ~1.2 hr$^{-1}$, which is comparable to ~0.9 hr$^{-1}$ employed in our simulations of the permeation process (Fig. 5d), and is consistent (at least in terms of order of magnitude) with literature data reported for a comparable system consisting of dithranol/liquid paraffin (~0.4 h$^{-1}$) [30]. Summarizing it appears, that the EC-modelling with the decay concept extension, along with the simplified forms of the model, provide decent tools for the permeation data analyses.

### 4.3 Two-beams DRPA and the interface sorption kinetics

In many practical cases involving mass transport between two phases, the role of the interface is reduced to that of a mathematical object: a null-thick boundary between subsystems of different characteristics (e.g. diffusion coefficient, mass transfer coefficient, etc.), where additional constraints or conditions like flux continuity are imposed on the transport equations. On the other hand, it is known that the physics of the interfaces is far more complex than a simple step-wise variation of physico-chemical properties, and includes thermodynamical definitions and concepts on the rheological effects and relaxation processes over a finite yet non-zero thick interfacial layer [31,32]. As such it is possible to differentiate between the bulk mass transport and the interfacial mass transfer rate influenced by both the interface equilibration kinetics and mass transfer coefficient. The transport parameters are related via the transport rate number: $N_R = \tau_{bulk}/\tau_{interface}$, where $\tau_{bulk}$ stand for the bulk diffusion characteristic time, while $\tau_{interface}$ represents characteristic timescales of the interface equilibration and interfacial mass transfer. The contributions of the bulk and interfacial transport can be indirectly divided and quantified by performing a broad PA depth-profiling studies accompanied with a proper mathematical analyses, as shown in [11]. In comparison, a direct access to the surface equilibration kinetics is provided by the time-dependent surface tension measurements, as described in [33]. Here, we propose a DRPA method for sensing interface optical properties (i.e. reflectivity) as an indicator of physicochemical state during the relaxation process.

By referring to the preliminary studies (see Fig. 2) it is expected that the membrane interface relaxations during the permeation process can be successfully monitored by a pair of 280 nm (membrane/air) and 530 nm (drug donor/membrane) light sources. The experimental results acquired with the use of the mentioned light sources are shown in Fig. 6. All the data were acquired during the same experiment. To prevent potential noise and interference in the experimental data, the multibeam scans were performed at two distinct modulation frequencies for each source fulfilling the modulation shift criterion.

For the 280 nm source (Fig. 6a) the most crucial observation concerns a discontinuity in the time-dependent DR magnitude run for $t = \tau_{disc}^{DR} \approx 0.4$ hr. The effect is noticeable for both modulation frequencies which rules out a single frequency disruption source, and is not observed in the experimental data for the 530 nm source. The observation can be explained upon the findings from our previous studies on similar system (pigment transport through DDC membrane of porosity ~50% and thickness of 30 μm) by means of PA depth-profiling for the bulk data analyses and wettability based techniques for the membrane/air interface processes quantification [27,33]. The bulk data analyses revealed the diffusion coefficient to be $D_{50\%} \approx 2.05 \cdot 10^{-9}$ $cm^2$ $s^{-1}$, again the value reasonably close to the one obtained here. On the other hand, the wettability data indicated no noticeable surface changes up to around 20 – 30 minutes of the experiment, despite the ongoing bulk diffusion ($\tau_{50\%} \approx 30$ min from the PA experiment). After the time period, an evolution in the surface characteristics was observed and lasted for another ~60 min up to the equilibration. The observations were eventually attributed to the relocation of the molecules from the sub-surface layer into surface layer and further chemical modifications including surface sorption. Bearing in mind the sequence of events, i.e. bulk diffusion (here $\tau \approx 0.3$ hr), membrane/air surface diffusion and the interface modifications, the discontinuity in the DR data for $\tau_{disc}^{DR}$ can be viewed as a rapid reflectivity enhancement originating from the chemical modification of the surface.

Beside the discontinuity effect, also a linear decay in the DR signal can be observed. By having in mind that the 280 nm DR originates from the light reflection from the membrane/air surface only, this is probably due to the light-induced degradation process of rather slow dynamics (10% per hour, as shown in Fig. 6a inset).

Fig. 6b shows the 530 nm DR data acquired for 37 and 120 Hz, which originate from the light reflected from both membrane interfaces. Both sets revealed similar exponential-like behaviour with the characteristic time of 0.17 hr. No effects of similar timescales were observed for the membrane/air interface and the discontinuity effect was not recognized in the 530 nm data (though $2\tau^{DR}$ remains relatively close to the $\tau_{disc}^{DR}$). The observations suggest that the signal originating from the donor/acceptor interface dominates that from the membrane/air interface. In fact, the donor sample consists of crystalline pigment structures dispersed in an ointment base, which substantially increases light scattering back to the detector. However, this scattering diminishes as the crystals dissolve in the presence of dodecanol—a permeable component of the membrane. Consequently, DR measurements at 530 nm monitor the evolution and equilibration of the interface state, driven by crystal dissolution and associated changes in surface reflectivity. By referring the transfer rate number obtained earlier for a somewhat similar system i.e. $N_R = 0.7$ for the membrane of porosity of 33%, and taking $\tau_{bulk} = \tau^{360}$ one expects the mass transfer rate limiting timescale of $\tau_{interface} \approx 1150\text{s} \approx 0.32$ hr, comparable to the $2\tau^{DR}$ value that is when the system reaches nearly equilibrium state.

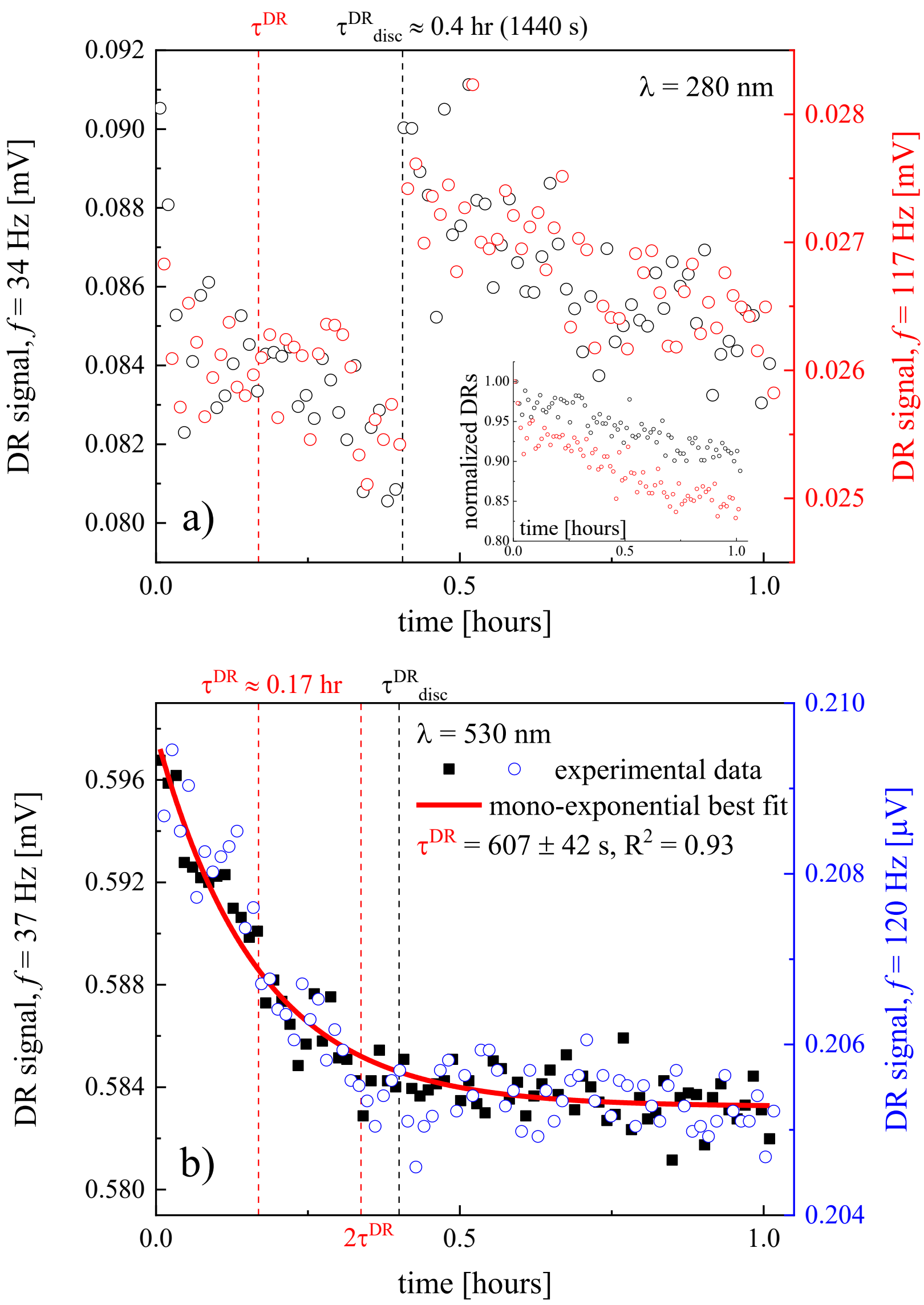

Fig. 6 DRPA results for the mass transport donor/acceptor system: a) probe light of 280 nm at 34 and 117 Hz for the air/membrane interface examination; the data exhibits discontinuity ($\tau_{disc}^{DR} \approx 0.4\ hr$) attributed to a rapid surface reflectivity change due to structural reorganization at the sample interface. The merged data run shows negative slope likely due to surface UV-induced degradation; b) probe light of 530 nm at 37 and 120 Hz for the membrane/ pigment donor interface examination. The data exhibits exponential-like decay related to the interface equilibration including crystalline pigment structure dissolution.

## 5 Conclusions

The paper elucidates the potential of the photoacoustics based techniques for a broad characterization of physical processes accompanying mass transport phenomena in thin films. Two base techniques: front-side detection PA and the diffuse reflection PA (DPRA), upgraded towards recently introduced multibeam setups and applied for the monitoring of a model photodegradable pigment permeation from a donor phase into a thin acceptor membrane, allowed for the multispecies bulk transport and interfacial kinetics quantification at the donor/acceptor and acceptor air interfaces simultaneously. The characterization of the methods for the membrane studies, along with the PA in the transmission mode serving as a supplementary tool for the adjusting experimental setup configuration, is summarized in Table 1.

Tab. 1 A summary of the photoacoustics-based modalities for the membrane transport studies.

| PA Mode | Application | Comments |
|---|---|---|
| Transmission (TR) | Initial monitoring of the samples optical characteristics and estimation of the light absorption coefficients for the needs of FSD and DR setups adjustements. | Works in a single detection mode that requires a sample change (i.e. the desired and reference materials).<br>As a spectral PA technique, the spectral source power needs to be taken into account. |
| *f*-domain FSD | Allows to track the amount of pigment concentration within a certain material depth of interest.<br>The MBPA mode extends the number of pigments to be examined.<br>If the frequency sweep is used, allows to record a distributed concentration evolutions. | Benefits from the TR data – works when $1/\alpha >$ sample thickness for the system in the equilibrium.<br>Requires to quantify thermal parameters of the samples prior the transport experiments (estimation of the cut-off frequency related to the sampling depth).<br>For the MBPA setups, a frequency separation condition needs to be applied. This implies a small deviation for the sampling depth of interest. |
| Diffuse reflection (DRPA) | Allows to track variations in the reflectivity of the interfaces as an indicators of the structural and chemical changes occurring during equilibration. | Benefits from the TR data – light sources from the optically transparent or opaque regions.<br>In the multibeam mode, both interfaces can be monitored simultaneously when the frequency separation condition is meet.<br>The signal is relatively weak and needs extended data sets for the kinetics analyses. |

The paper also employs quantification of sorption data using an electrical circuit-based model of diffusion-reaction transport. This model was specifically designed for degradable trace molecules following a single degradation pathway to a stable photoproduct, with mass transport occurring from a donor to an acceptor compartment. System dynamics are captured by an evolution matrix that couples the two molecular populations through decay rates and mass transfer coefficients—driven by concentration gradients and linked to diffusion coefficients via electrical circuit (EC) resistivities. A simplified version of the model, tailored to the experimental conditions and valid for short times, was introduced and validated against data acquired via the membrane-based photoacoustic (MBPA) method. Although the mathematical expressions governing the concentration evolution of the primary pigment and its photoproduct in the membrane were deliberately simple, the diffusion coefficients extracted from data-fitting exhibited promising agreement with predictions from the general model, particularly in terms of order of magnitude. Thus, the study establishes a robust, extensible modeling framework for quantitative analysis of mass transfer processes in applied physics contexts.